# A low Complexity Wireless Gigabit Ethernet IFoF 60 GHz H/W Platform and Issues

L.Rakotondrainibe[1], I. Siaud[2]. Y. Kokar[1], G. Zaharia[1], F. Brunet[2], E. Tanguy[3], G. El Zein[1]
[1] IETR-INSA, UMR CNRS 6164, 20, Rennes, France
[2] FTR&D-Orange Labs, Cesson Sévigné, Rennes, France
[3] Université de Nantes, IREENA, Nantes, France

***Abstract*—This paper proposes a complete IFoF system architecture derived from simplified IEEE802.15.3c PHY layer proposal to successfully ensure near 1 Gbps on the air interface. The system architecture utilizes low complexity baseband processing modules. The byte/frame synchronization technique is designed to provide a high value of preamble detection probability and a very small value of the false detection probability. Conventional Reed-Solomon RS (255, 239) coding is used for Channel Forward Error Correction (FEC). Good communication link quality and Bit Error Rate (BER) results at 875 Mbps are achieved with directional antennas.**

## I. INTRODUCTION

DEMANDS for high-speed multimedia data communications, such as a high speed file transfer and real-time video streaming, are markedly increasing. The 60 GHz band, due to a large bandwidth is one of the most promising solutions to achieve a gigabit class for short distance high speed communications.

This paper presents a hardware IFoF system architecture dedicated to 60 GHz Ultra-Wide Band (UWB) radio transmission for Wireless Personal Area Network (WPAN) applications into a Point-to-MultiPoint (P2MP) deployment manner. The work has been developed within the framework of a French consortium, the Techim@ges project [1] included in "Media & Network" Cluster. The hardware platform is oriented on a low complexity implementation and a backward compatibility with IEEE802.15.3c/ECMA-387 PHY layer proposals [2][3] by considering similar 2 GHz channelization in {57, 64} GHz RF band defined by FCC 47 CFR 15.255. IEEE802.15.3c standardization WG is currently defining a Multi Gigabit Wireless System (MGWS) in the 60 GHz RF band for WPAN applications [2]. The proposed IFoF architecture system is envisioned as low cost IFoF architecture for future IEE802.15.3c and ECMA-387 [3][4] proposal evolutions enable to encounter Remote Antenna Unit (RAU) and extend the radio coverage of millimeter waves (Mm-wave) P2MP (Point to Multi-Point) WPAN systems. Mm-Wave Radio over Fibre (RoF) transmissions for WPAN applications are designed in the IST- IPHOBAC project [5] with efforts concentred on optical components and advanced mm-wave signal generation using heterodyning techniques and photonic components. The IPHOBAC radio component is concentrated on advanced UWB-OFDM (Orthogonal Frequency Division Multiplexing) baseband processing [6] and 60 GHz transposed Wi-Media radio signal [7]. The Radiofrequency (RF) UWB signal is converted to an optical signal and transported through the fibre, followed with photo-detection and radio emission. In Techim@ges project, IFoF architecture systems are investigated with single carrier radio transmission combined with Frequency Domain Equalization (FDE).

After an introduction of MGWS context and standardisation progress, a description of the WPAN 60 GHz propagation channel is carried out on the basis of experimental measurements lead in the IST-FP6-MAGNET project. This characterization clearly justifies the mm-wave radio coverage limitations due to 60 GHz multipath signature in the face of MGWS system and path loss attenuation.

The second part of the paper describes the H/W IFoF architecture system developed in Techim@ges and associated performance in the case of Point-to-Point (P2P) transmission. The system uses the same IEEE802.15.3c channelization in order to ensure coexistence between systems and easily evolve the proposed architecture to IEEE802.15.3c proposal.

## II. STANDARDIZATION BACKGROUND

Applications such as high speed file transfer and streaming video in a point-to-point link configuration are considered in use cases defined in IEEE802.15.3c WG. Dedicated usage models have been defined to cover data rate ranged from 50 Mbps to 6-7 Gbps. For that purpose, a dedicated channelization based on 2 GHz channel bandwidth has been defined in the {57-64} GHz. The channelization is common for single carrier and OFDM transmissions as illustrated on the Fig.1.

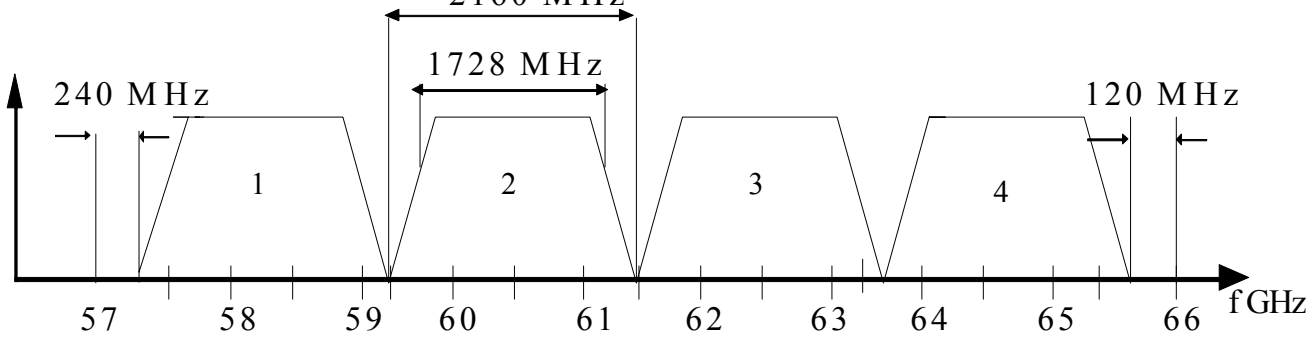

Fig. 1. 60 GHz regulations issued from FCC 47 CFR 15.255

If single carrier transmissions modes are split into several classes depending on data rate range. Simple FEC coding and modulation are considered for Low Data Rate (LDR) configurations using Reed-Solomon code as FEC scheme, as illustrated on the table I. A FEC scheme is determined with the trade-off between higher coding gain and hardware complexity.

The ECMA-387 Standard edited in December 2008 by the ECMA TC-48 group specifies a physical layer (PHY), distributed medium access control (MAC) sub-layer, and an High Definition Multimedia Interface (HDMI) protocol adaptation layer (PAL) for 60 GHz wireless networks. Physical (PHY) layer proposal is derived from IEEE802.15.c proposal with a splitting of proposals in three modes A, B and C [3].

TABLE I
60 GHz SINGLE CARRIER MODULATION SCHEME (IEEE 802.15.3 C[2])

| MCS class | MCS Identifier | PHY-SAP rate Mbps | Modulation scheme | Spreading factor | FEC TYPE | FEC rate |
|---|---|---|---|---|---|---|
| Class 1 | LR1 | 50.6(CR)/379.6/ 759.2/1518.4 (MLR) | π/2- BPSK/ (G)MSK | 32/4/2/1 | RS(255,239) | 0.937 |
| | LR2 | 607.5/1215.0 | π/2- BPSK/ (G)MSK | 2/1 | LDPC(576,432) | 0.750 |
| | LR3 | 810.0 | π/2- BPSK/ (G)MSK | 1 | LDPC(576,288) | 0.500 |
| Class 2 | MR1 | 162.0 | π/2-QPSK | 1 | LDPC(576,288) | 0.500 |
| | MR2 | 2430.0 | π/2-QPSK | 1 | LDPC(576,432) | 0.750 |
| | MR3 | 2835.0 | π/2-QPSK | 1 | LDPC(576,504) | 0.875 |
| | MR4 | 3024.0 | π/2-QPSK | 1 | LDPC(1440,1344) | 0.933 |
| | MR5 | 30356.7 | π/2-QPSK | 1 | RS(255,239) | 0.937 |

## III. 60 GHz PROPAGATION SIGNATURE

The 60 GHz propagation signature issued from Orange Labs measurements and models are detailed in the recent Wiley Book Edition [9].

60 GHz multipath signature is characterized with relative high delay spread values relatively to MGWS symbol and sampling rates. The Average power delay profiles (APDP) are compliant with multi-cluster model derived from the Saleh-Valenzuela model defined in [10] and experimental results. APDP are characterized with several clusters having significant relative amplitude involving Root Mean Square (RMS) delay spread up to 15-20 ns. RMS delay spread for both Orange labs measurements and CEPD model realizations and IEEE802.15.3c are summarized in tables II and III.

TABLE II
60 GHz MULTIPATH SIGNATURE OF ORANGE LABS MULTIPATH MODELS [9]

| | CEPD LOS (Tx-72°, Rx60°) | CEPD NLOS (Tx-72°,Rx-60°) | CEPD critical NLOS (Tx-72°, Rx-60°) |
|---|---|---|---|
| Distance | 7.615 m | 8.36 m | 12.1 m |
| RMS delay spread $\sigma_{DS}$ | 2.73 ns | 7.38 ns | 12.70 ns |
| $\sigma_{DS}$ Standard Deviation | 1.36 dB | 1.97 dB | 0.8 dB |
| Power Standard Deviation | 0.66 dB | 1.97 dB | 0.8 dB |
| Coherence bandwidth ρ=0.5 | 156 MHz | 58 MHz | 11 MHz |
| Time Coherence 0.5 | 36 ms | 8 ms | 4 ms |

Path-loss models are derived from the modified Friis transmission equation including exponential distance dependency and standard deviation of measurement points. Path-loss models are described by the equation:

$$PL_{MFS}\left(d\right)=PL_{FS}\left(d_0\right)+10\cdot n\,log_{10}\left(\frac{d}{d_0}\right)+\sigma$$

$PL_{FS}(d)$ is the free space path-loss model; σ is the standard deviation of experimental attenuation points used to generate the model.

TABLE III
PROPAGATION PARAMETERS OF 60 GHZ WPAN MODIFIED SALEH-VALENZUELA MODEL FROM NICT

| Parameter | Office Value | Desktop Value | Distribution |
|---|---|---|---|
| Cluster decay factor | 19.44 ns | 4.01 ns | Exponential |
| Ray decay factor | 0.42 ns | 0.58 ns | Exponential |
| Cluster gain standard deviation | 1.82 ns | 2.70 ns | Lognormal |
| Ray gain standard deviation | 1.88 ns | 1.90 ns | Lognormal |
| RMS delay spread | 24.82 ns | 2.27 ns | |
| Average number of clusters | 6 | 3 | uniform |

A selection of measurements has been carried out upon measurements to generate accurate path-loss models with limited path-loss dispersion. Selected path-loss models exhibiting a standard deviation inferior to 6 dB are preferred. These models are extracted from [9]. Fig. 2 shows path loss models for LOS and NLOS configurations.

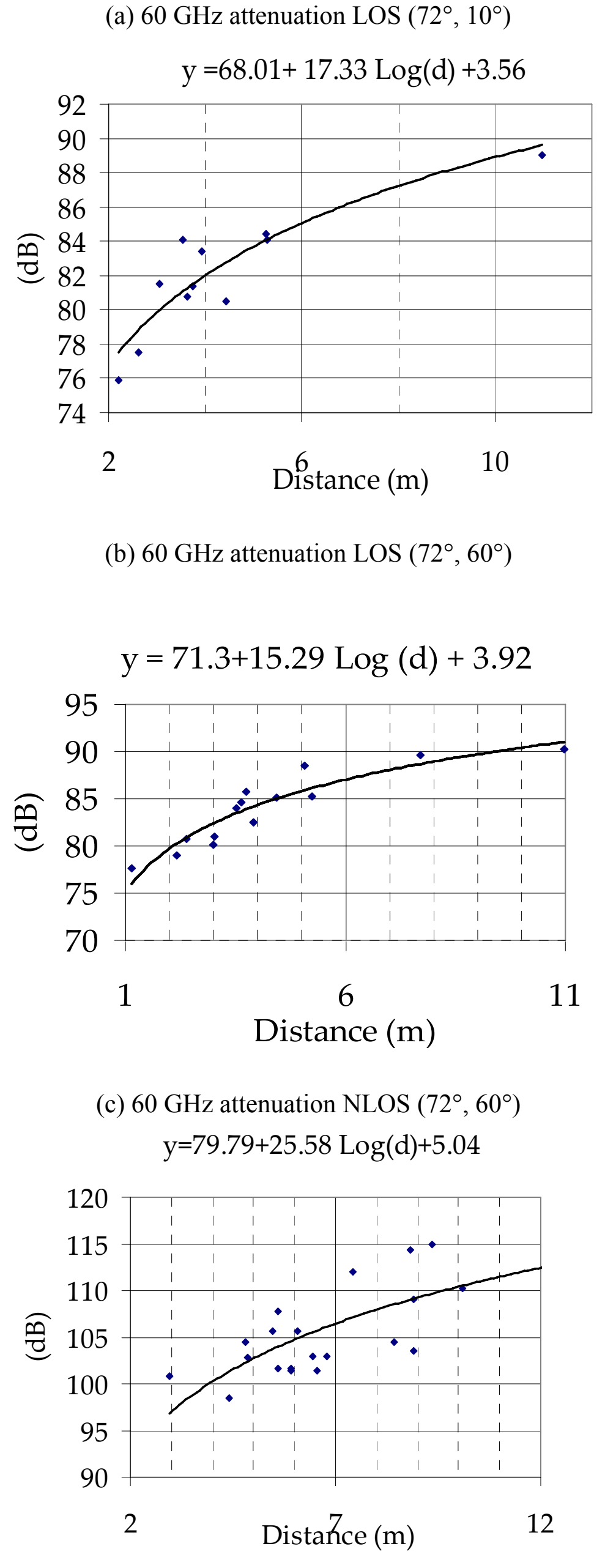

Fig. 2. Path los model Orange Labs: (a), (b) recommended, (c) high dispersive model

## IV. IFoF SYSTEM ARCHITECTURE AND BASEBAND DSP DESCRIPTION

The Gigabit-Ethernet interface is used to connect a home server to a wireless link with around 800 Mbps bit rate. Fig.3 and Fig.4 show the transmitter and receiver Gigabit Ethernet interface. The Gigabit Media Independent Interface (GMII) is an interface between the Media Access Control (MAC) device and the physical layer (PHY). The interface defines speeds up to 1 Gbps, implemented using 8 bits data interface clocked synchronized at 125 MHz. However, this frequency is different from the clock sampling data (104.54 MHz) generated by a clock manager, as explained latter. These clock synchronisms involve data packet loss. In order to avoid jitter transmission, a programmable circuit (FPGA) is used as a part of buffer memory.

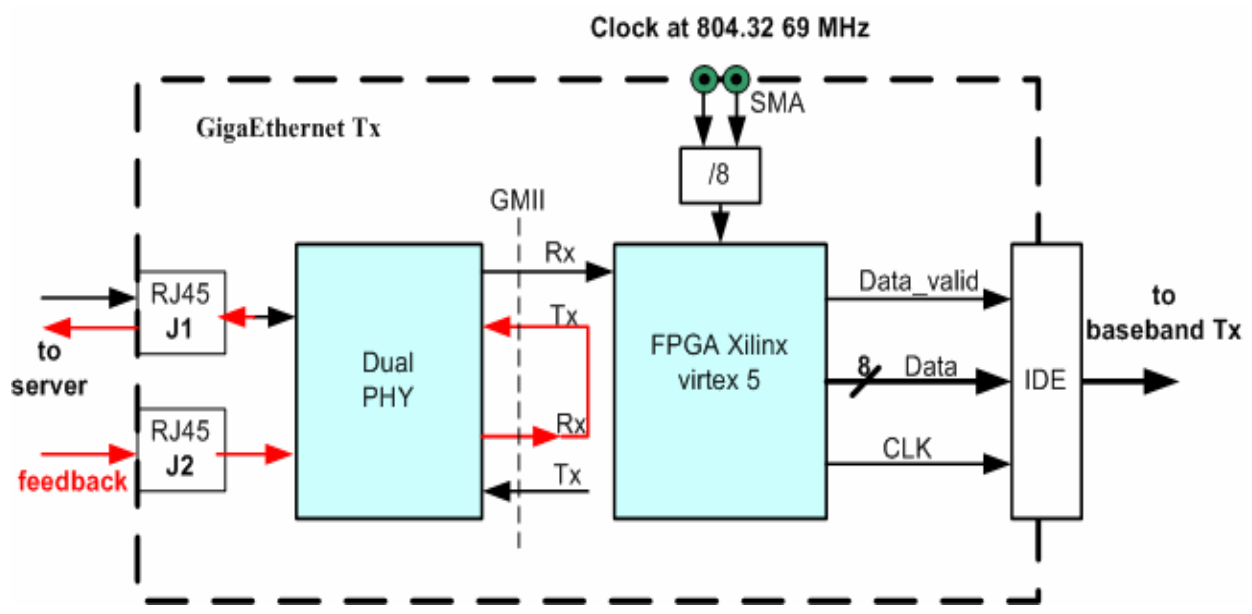

Fig. 3. Transmitter Gigabit Ethernet interface

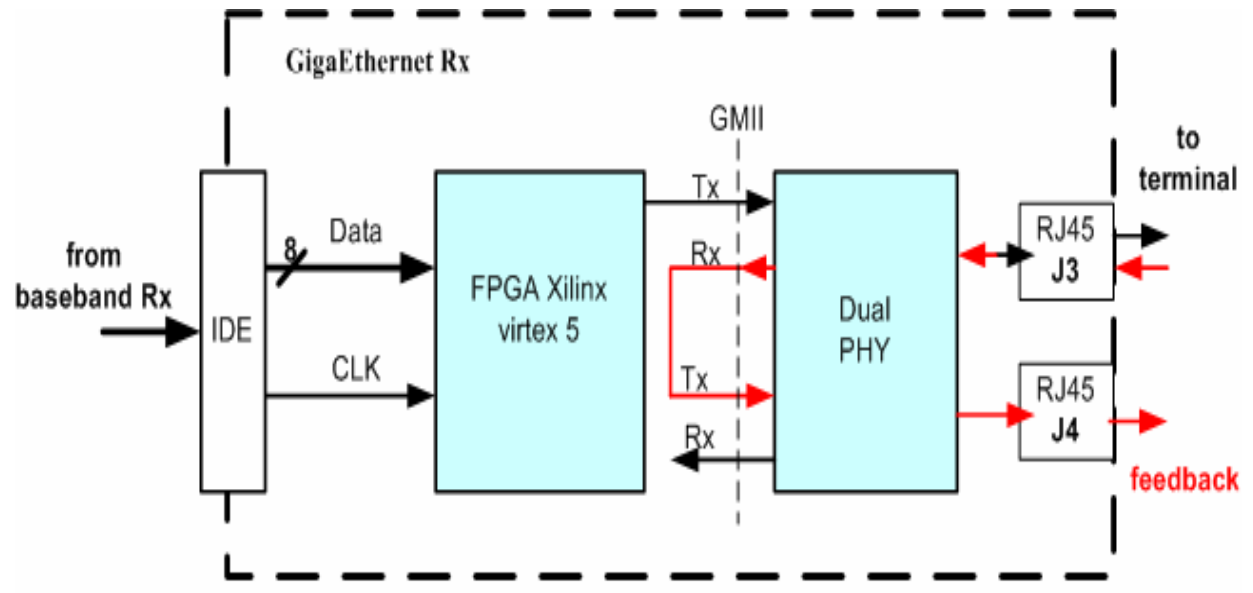

Fig. 4. Receiver Gigabit Ethernet interface

The IFoF system uses a Single Carrier (SC) radio transmission based on a differential encoded binary phase shift keying (DBPSK) modulation and a non-coherent demodulation [14]. The Reed-Solomon (RS) coding are used for channel forward error correction (FEC) to ensure compatibility with IEEE802.15.3c SC modes. The general link design of the system is shown in Fig. 5 and Fig. 6.

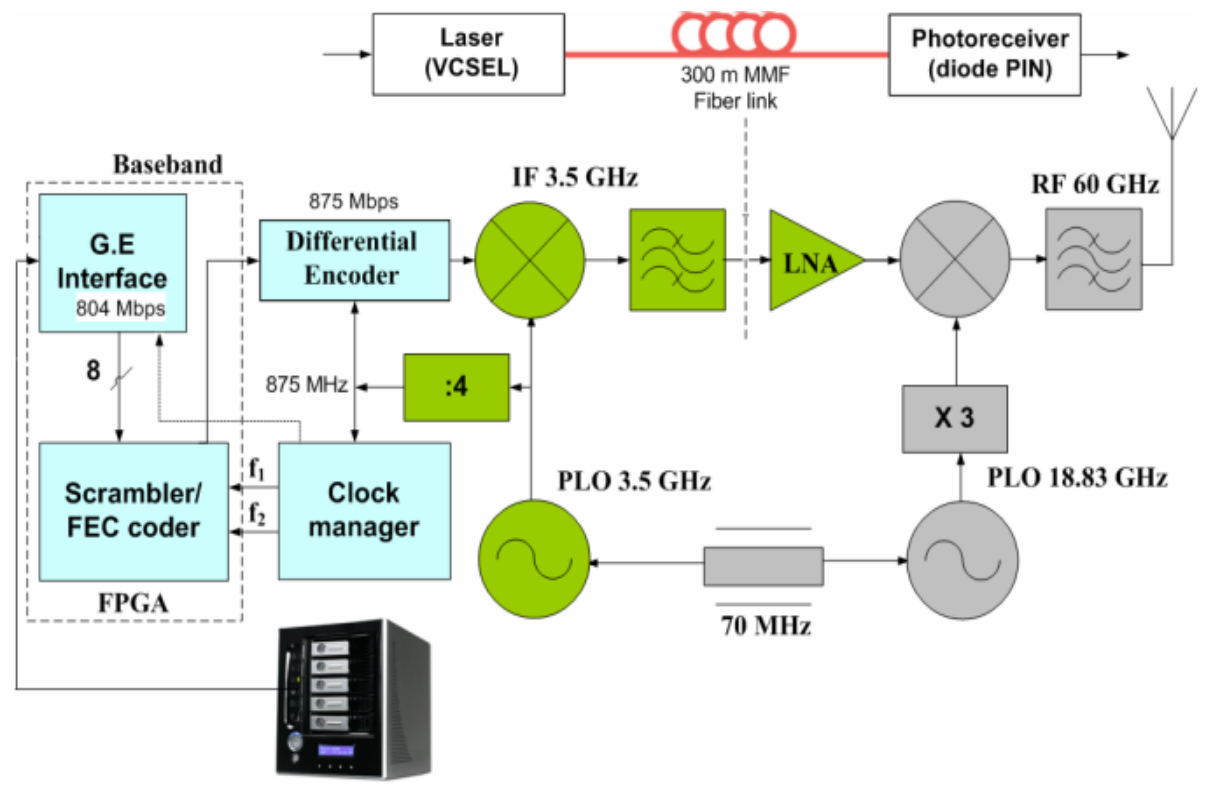

Fig. 5. 60 GHz Wireless Gigabit Ethernet transmitter

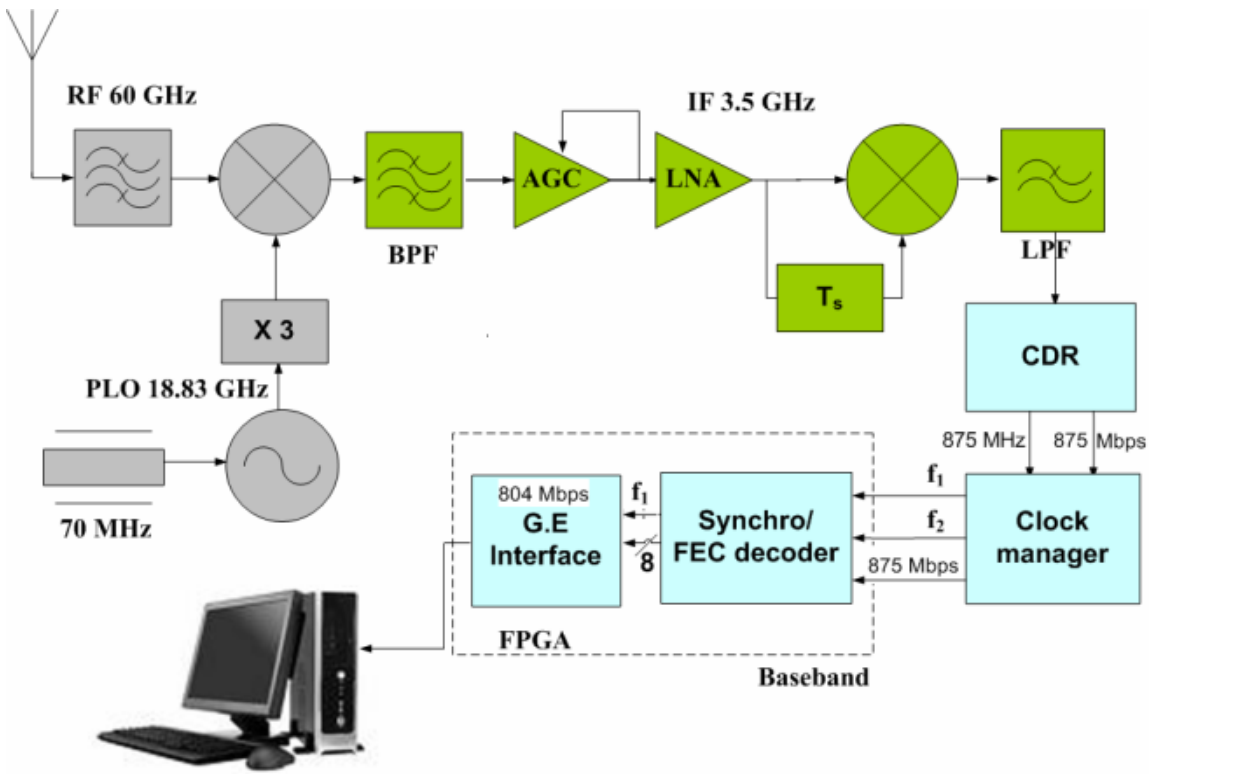

Fig. 6. 60 GHz Wireless Gigabit Ethernet receiver

DBPSK modulation is used to avoid the phase ambiguity at the receiver. Compared to higher order constellations or multi-carrier modulation (OFDM), this system is more resistant to phase noise and power amplifiers (PAs) non-linearities. OFDM requires a 6 dB back-off for PAs in the transmission, high stability and low phase noise for local oscillators. On the other hand, enhancements on the baseband processing compensates 6 dB back-off [6].

### A. *Intermediate and radiofrequency architecteture*

After the channel coding and the scrambling, the input data are differentially encoded using logic circuits (PECL). The differential encoder performs the delayed modulo two addition of the input data bit with the output bit. The obtained data modulate an Intermediate Frequency (IF) carrier generated by a 3.5 GHz phase locked oscillator (PLO) with a 70 MHz external reference. The IF signal is fed into a band-pass filter (BPF) with 2 GHz bandwidth, and then transmitted through a 300 meters fibre link. This IF signal is used to modulate directly the current of the Vertical Cavity Surface Emitting Laser (VCSEL) operating at 850 nm through a band-pass filter with a bandwidth of 2 GHz. The VCSEL input RF power must not exceed -3 dBm to avoid signal distortions. After transmission, the optical signal is converted to an electrical signal by a PIN diode and amplified. The overall RoF link has 0 dB gain and an 8 GHz bandwidth. This bandwidth or the fibre length could be increased if necessary by using a VCSEL and a photodetector of broader bandwidth.

Following the RoF link, the IF signal is sent to the RF block. This block is composed of a mixer, a frequency tripler, a PLO at 18.83 GHz and a band-pass filter (59-61 GHz). The local oscillator frequency is obtained with an 18.83 GHz PLO with the same 70 MHz reference and a frequency tripler. The phase noise of the 18.83 GHz PLO signal is about –110 dBc/Hz at 10 kHz off-carrier. The BPF prevents spill-over into adjacent channels and removes out-of-band spurious signals caused by the modulator operation. The 0 dBm obtained signal is fed into the horn antenna with a gain of 22.4 dBi and a half-power beamwidth (HPBW) of 12°.

The input band-pass filter removes the out-of-band noise and adjacent channel interference. The RF signal at the output of the filter is down-converted to an IF signal centered at 3.5 GHz which is fed into a band-pass filter with a bandwidth of 2 GHz. An Automatic Gain Control (AGC), with a dynamic range of 20 dB, is used to ensure a quasi-constant signal level at the demodulator input. A Low Noise

Amplifier (LNA) with a gain of 40 dB is used to achieve sufficient gain. A simple differential demodulation enables the coded signal to be demodulated and decoded. Compared to a coherent demodulation, this method is less performing in additive white Gaussian noise (AWGN) channel. The significant impact on the system caused by the radio channel is the frequency selectivity which induces inter-symbol interference (ISI). Differential demodulation is more resistant to multi-path interference effects. Indeed, the differential demodulation, based on a mixer and a delay line (Ts = 1.14 ns), compares the signal phase of two consecutive symbols. Due to the product of two consecutive symbols, the rate between the main lobe and the second lobe of the impulse response of the channel is increased. Following the loop, a Low-Pass Filter (LPF) with 1 GHz cut-off frequency removes high-frequency components of the obtained signal. For the Clock and Data Recovery (CDR) circuit, long sequences of '0' or '1' must be avoided. In this case, the use of a scrambler at the transmitter and a descrambler at the receiver is essential.

### *B. Baseband architecture*

Baseband blocks are implemented in programmable circuit FPGA Xilinx Virtex 4. The transmitter baseband (BB-Tx) is composed of the channel coding, the frame formatting block and the scrambler, as shown in Fig. 7. The channel coding is realized with the RS (255, 239) coder, using byte data. The frame format consists of 4 preamble bytes, 239 data burst bytes, 1 frame header byte and 16 check bytes. To achieve the frame synchronization in the packet-based communication system, a known sequence (preamble) is sent at the beginning of each packet. The used preamble is a Pseudo-Noise (PN) sequence of 31 bits + 1 bit to provide 4 preamble bytes. One more byte called frame header should be added besides 4 preamble bytes to obtain a multiple of 4 coded data length useful for the scrambling operation. Hence, owing to the frame structure, two different clock frequencies $f_1$ and $f_2$ are used in the baseband block:

$$f_1 = \frac{F_1}{8} = 100.54\ \text{MHz}, \quad f_2 = \frac{F_2}{8} = 109.37\ \text{MHz}.$$

where: $F_2 = \frac{IF}{4} = 875\ \text{MHz}$ and $\frac{F_1}{F_2} = \frac{239}{260}$.

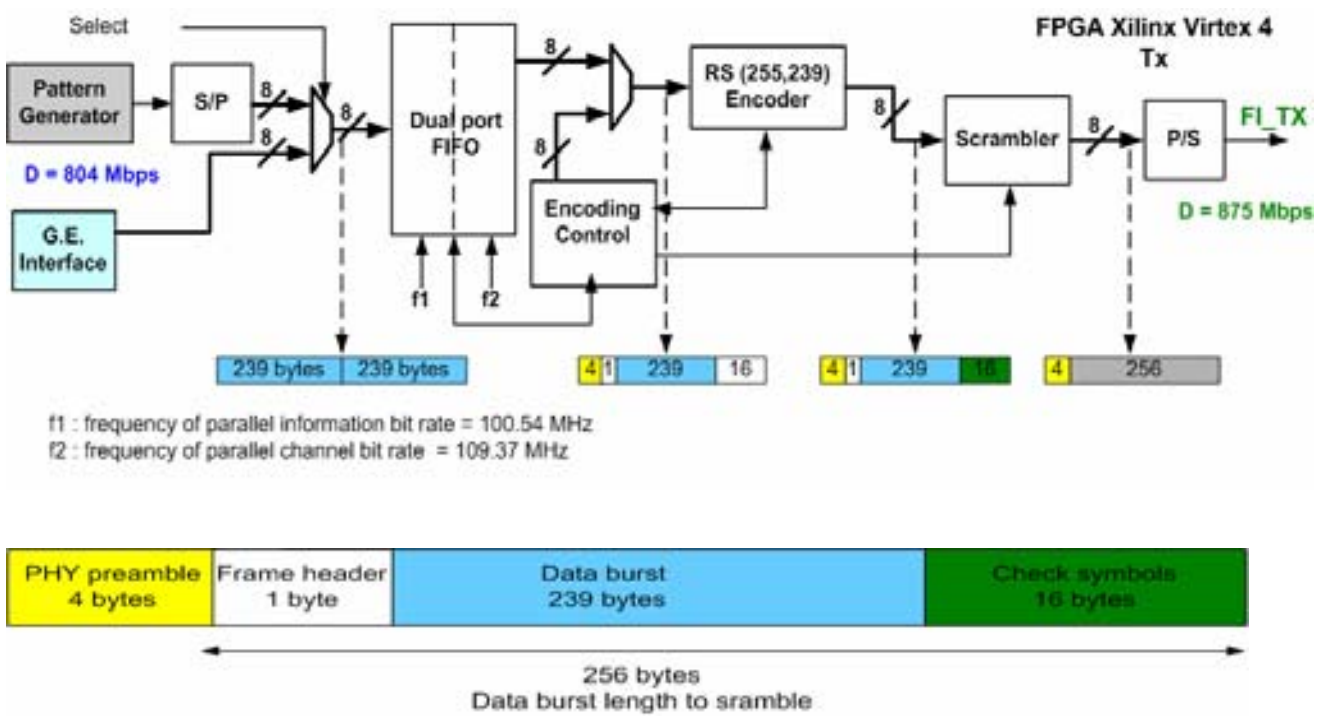

Fig.7. Transmitter baseband architecture and frame structure

The structure of the frame is generated as follows: byte streams are written into the FIFO dual port at $f_1$. The FIFO dual port has been set up to use two different frequencies for writing at $f_1$ and reading at $f_2$. Therefore, reading can be started when the FIFO is half-full. The encoding control is a module that monitors the RS encoder and the scrambler. A FEC scheme is determined with the trade-off between higher coding gain and hardware complexity. The RS encoder reads one byte every clock period. After 239 clocks periods, the encoding control interrupts the bytes transfer during 21 triggered clock periods. During this interruption, 4 preamble bytes are added; the encoder takes 239 data bytes and appends 16 control bytes to make a code word of 255 bytes. The encoding control adds one more byte for the balanced scrambling operation so that the number of 256 data coded bytes is a multiple of 4. The scrambler is a PN-sequence of 31 bits + 1 bit to obtain 4 bytes. This scrambler is chosen in order to provide the lowest cross-correlation values between the received data and the 4 bytes PHY preamble. This method reduces the number of false detection of the PHY preamble within the scrambled data. The received byte stream is finally parallel-to-serial (P/S) converted just before the differential encoder.

Fig. 8 shows the receiver baseband architecture. The recovered data from CDR at 875 Mbps are S/P converted into a byte frame.

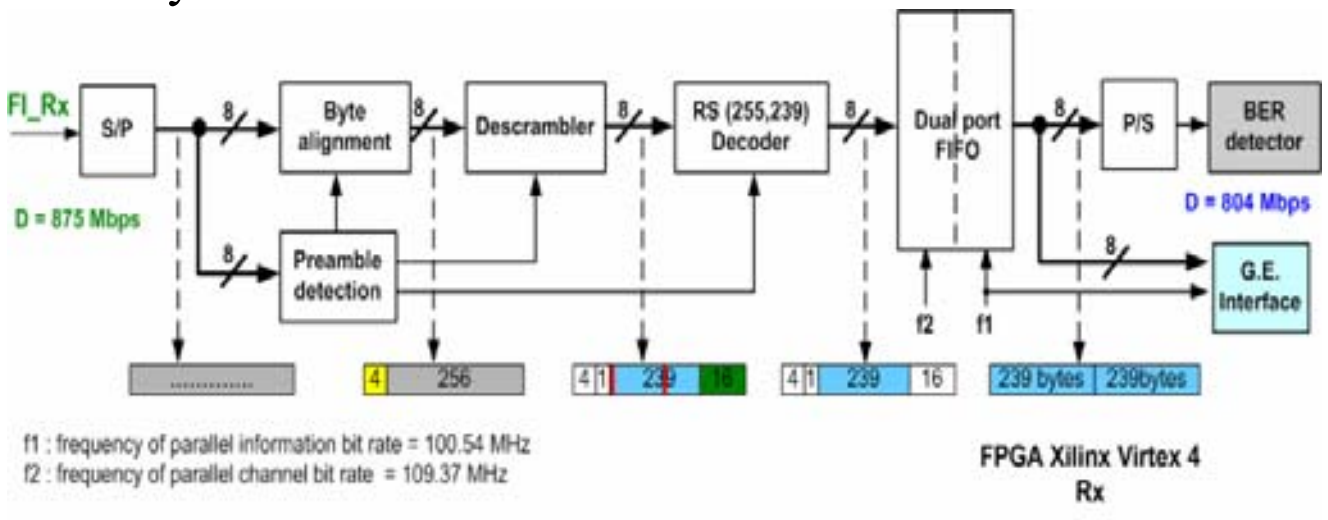

Fig. 8. Receiver baseband architecture

Fig. 9 shows the method used for the preamble detection architecture and the byte alignment.

The preamble detection is based on the cross-correlation between 32 successive received bits and the internal 32 bits PHY preamble. In addition, each correlator must analyze a 1-bit shifted sequence. Hence, the preamble detection is performed with 32 + 7 = 39 bits (+ 7 because of different possible shifts of a byte). In all, there are 8 corrrelators of 32 bits within each correlator-bank, as shown in Fig. 9.

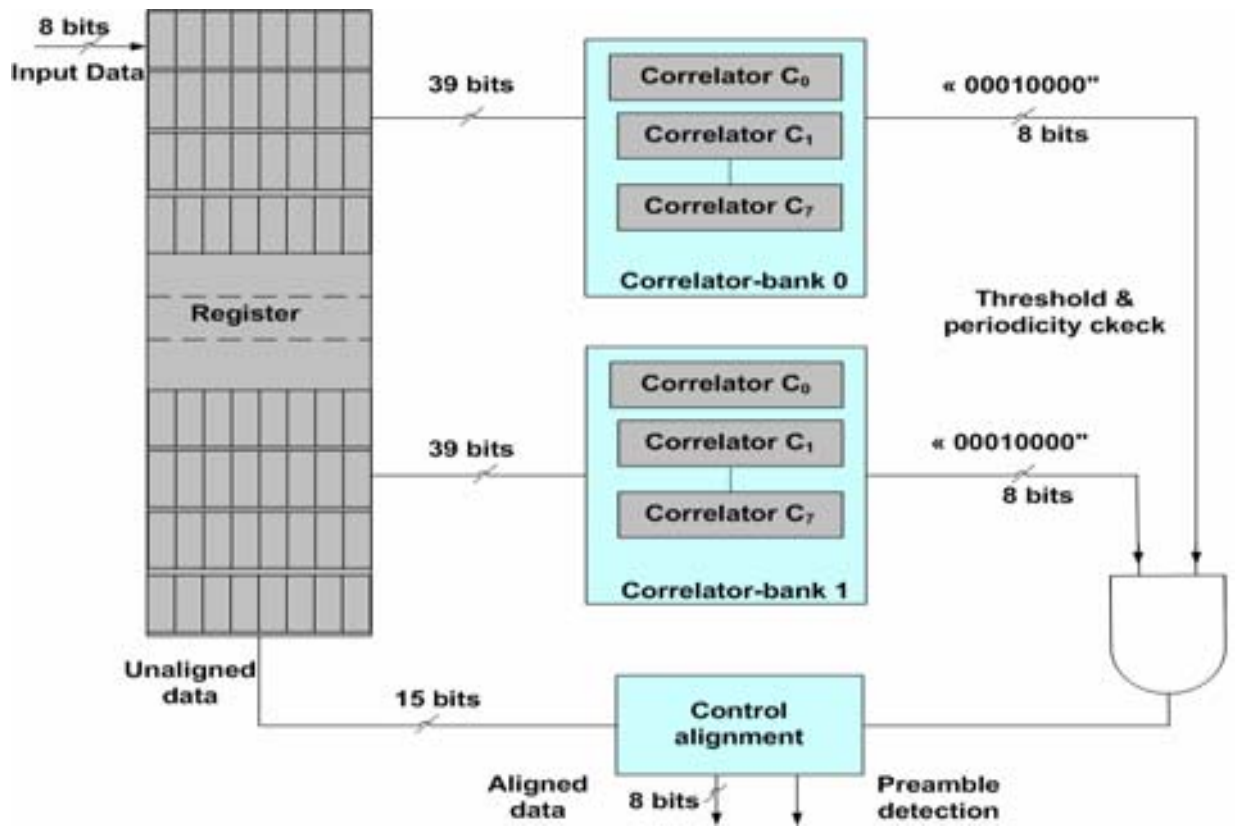

Fig. 9. Preamble detection with periodicity and control alignment

Each value of the correlation between the preamble and the received sequence data is compared to a threshold. The Frame synchronization performance is characterized by the miss detection probability and false alarm probability to be determined. This threshold is chosen in order to obtain the best trade-off between a high value of the detection probability and a very small value of the miss detection probability. The decision is performed from 264 successive bytes: preamble1 + data1 (256 bytes) + preamble2 stored in a register. If the preamble detection is indicated in each correlator-bank by the same correlator $C_k$, the operation is validated.

On the other hand, the byte alignment is established with 8 + 7 = 15 successive bits (a 1-bit shifted sequence).

After the byte alignment and the preamble detection, the descrambler performs the addition modulo 2 between 4 successive received data bytes and the 4 bytes scrambler sequence. The RS decoder processes the descrambled bytes and attempts to correct the errors. The RS (255, 239) decoder can correct up to 8 erroneous bytes and operates at a high clock frequency (109.37 MHz). The byte streams obtained at the decoder output are written discontinuously in the FIFO dual port memory at the clock frequency $f_2$. The other frequency 100.54 MHz read out continuously the data bytes stored in FIFO memory. Then, the byte streams are finally transmitted to the Gigabit Ethernet Interface, as depicted in Fig.4.

## V. IFoF System Architecture and Performance

A vector network analyzer (HP 8753D) was used to determine the frequency response and the impulse response of the both RF blocks (Tx + Rx) including LOS channel. The goal was to measure the system bandwidth and to estimate the effects of the multipath channel. The measurements were realised in an empty corridor with approximate dimensions of 15x2.4x3.1 $m^3$ where the major part of the transmitted power is focused in the direction of the direct path. The RF-Tx and RF-Rx were placed at a height of 1.5 m. After calibrating the measurement set-up, a frequency response of 2 GHz bandwidth is achieved, as shown in Fig. 10.

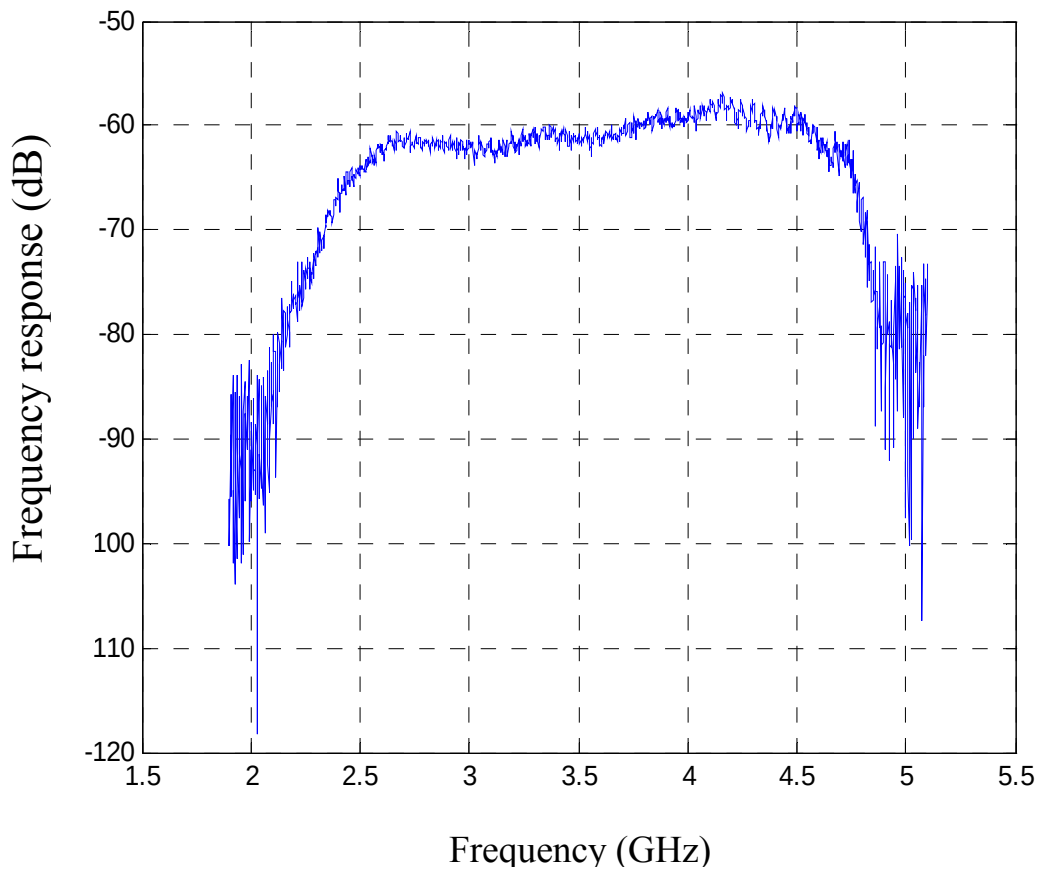

Fig. 10. Frequency response of RF (Tx-Rx) blocks)

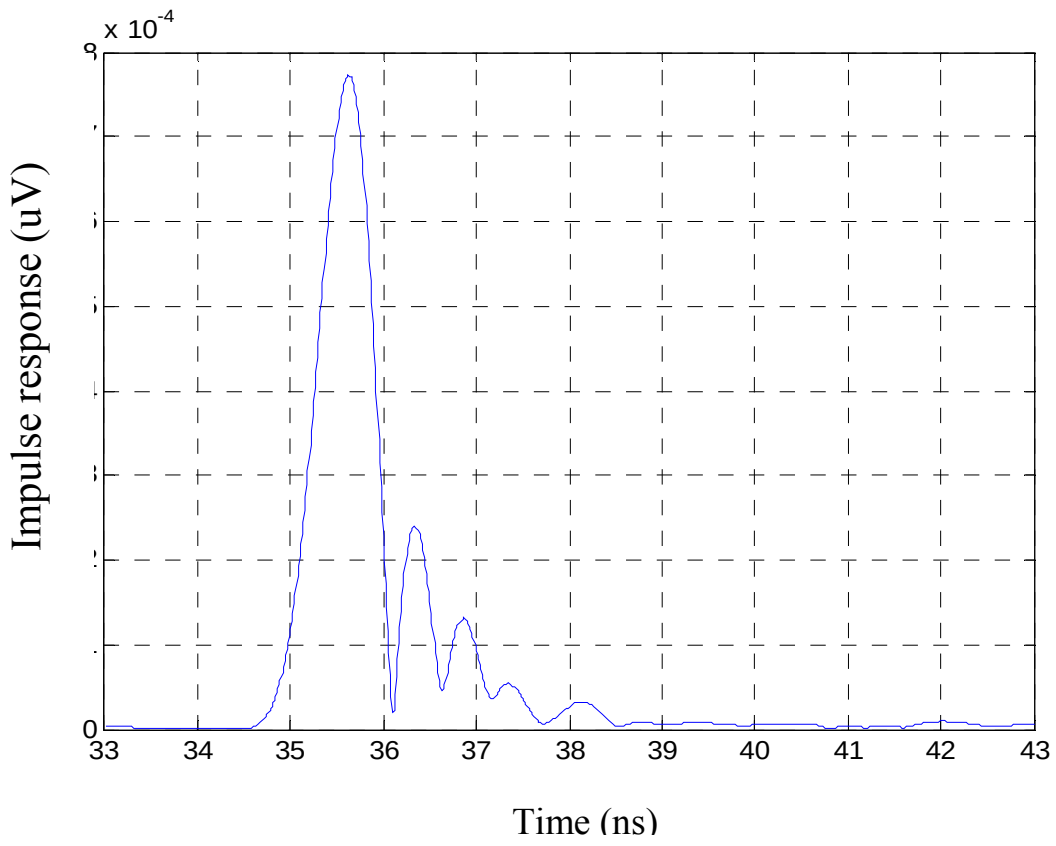

Fig. 11. Impulse response of RF (Tx-Rx) blocks)

As shown in Fig. 11, the RF blocks with their antennas placed at 10 m Tx-Rx distance present an impulse response with several side lobes which are mainly due to RF components imperfections.

Moreover, in order to evaluate the transmission performance, we used a HP70841B pattern generator to transmit PN sequences and a HP 708842B error detector at the receiver. Figure 15 indicates a high quality of the received data at 875 Mbps rate using an oscilloscope with 2 GHz bandwidth.

In our experiment, four antennas were used: two horns and two patch antennas. Each horn antenna has 22.4 dBi gain and 12° HPBW and each patch antenna has 8 dBi gain and 30° HPBW. 8 and Fig. 13 shows the BER performance at 875 Mbps in terms of the distance Tx-Rx.

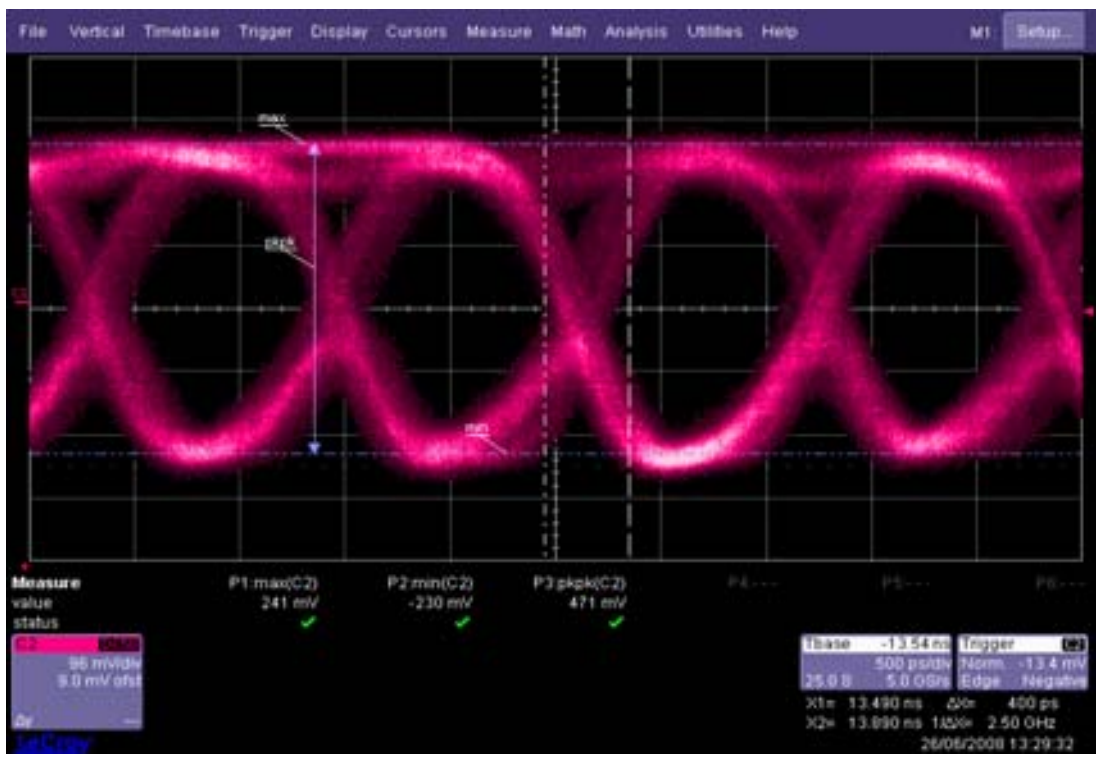

Fig. 12. Eye diagram at 875 Mbps, 10 m Tx-Rx distance

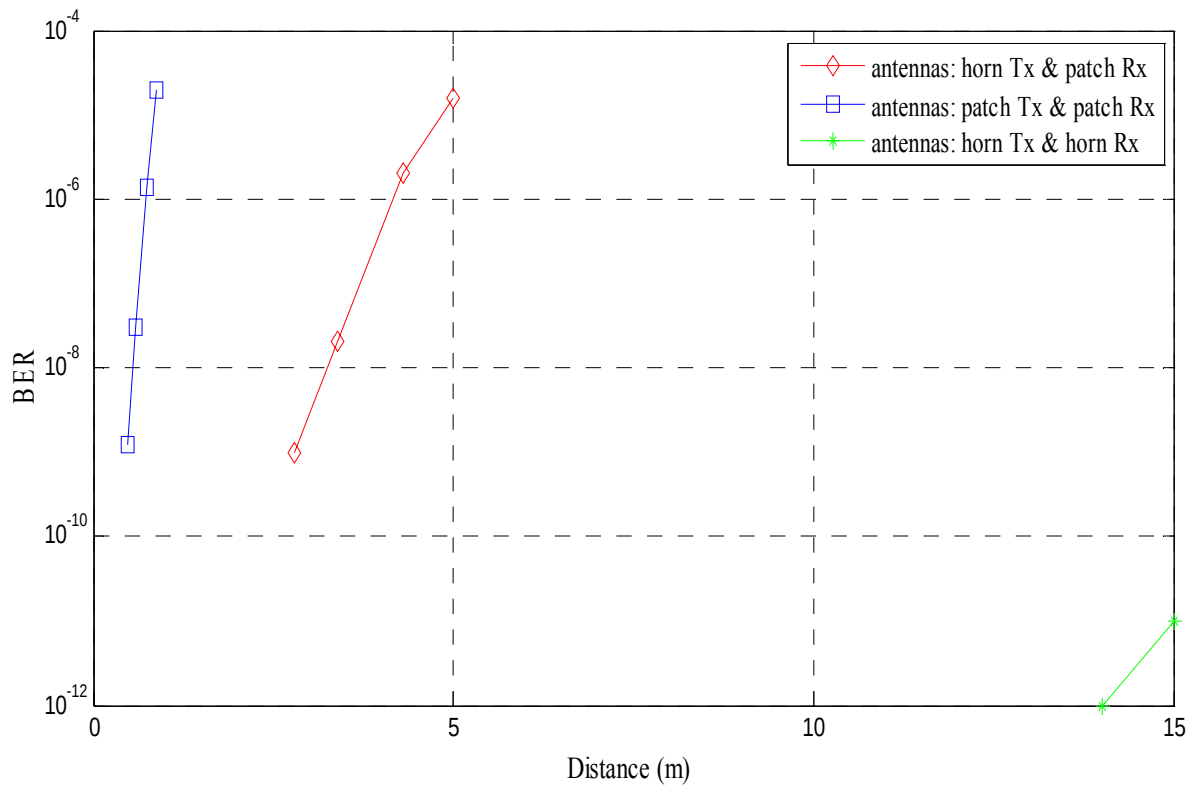

Fig. 13. BER performance at 875 Mbps (without channel coding)

From the measurement results, the use of high-directivity horn antennas gives a remarkable BER performance and greatly reduces the harmful effects of multipath propagation. System applications such as data transfer in large rooms can be obtained if high gain antennas are used. However, the Tx-Rx antennas have to be properly aligned; otherwise the beam-pointing errors will cause an important reduction of the communication quality due to an increased BER. For directive configurations, if the direct path is blocked by moving objects, the communication can be completely lost.

The system using high gain antennas is acceptable for point-to-point communications links, with minimal multi-path interference. However, if antennas beamwidth is large, equalization should be used to overcome multi-path interference while maintaining a high data rate. If patch antennas are used, more gain is desired for RF front-ends to compensate the link budget. Future work will provide the system performance results with channel coding.

## VI. Conclusion

This paper presents the design and the implementation of a 60 GHz communication system for WPAN applications in point-to-point or point-to-multipoint configurations. The proposed system provides a good trade-off between performance and complexity. An original method used for the byte synchronization is also described. This method allows a high preamble detection probability and a very small false detection. For a reliable communication at data rates around 1 Gbps over distances up to 10 m, antennas must have a relatively high gain.

We planned to increase the data rate using higher order modulations. Equalization methods are still under study. The demonstrator will be further enhanced to prove the feasibility of wireless communications at data rates of several Gbps in different configurations, especially in the case of non line-of-sight (NLOS) environments.

## Acknowledgment

Part of this work has been supported by the "Région Bretagne" in Comidom project. We are grateful to Yvon Dutertre (Techim@ges manager) and Ali Louzir, and the whole 60 GHz-RF design group for useful discussion. The authors would like to acknowledge Guy Grunfelder (CNRS engineer) for his contributions to the system realization.

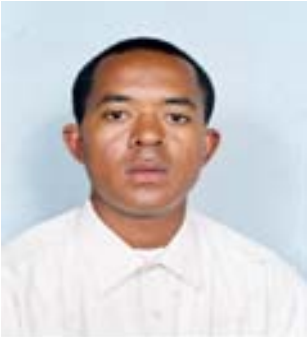

**Lahatra Rakotondrainibe** was graduated in telecommunications engineering at Polytechnic Superior School of Antananarivo, Madagascar in 2005. In 2007, he graduated in master research of "Microtechnologies, Architectures, Networks and Communication systems" from the National Institute for the Applied Science (INSA of Rennes), France. Since October 2007, he is a PhD student at IETR-INSA in electronics and communications systems. He has been involved within the framework of a French consortium, the Techim@ges and Comidom projects concerning the realization of high data rate 60 GHz technologies for multimedia applications.

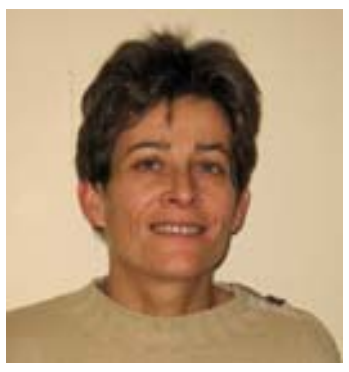

**Isabelle Siaud** received the Electronic Master Diploma from the University (UMPC Paris VI) in 1992. From 1993 to 1998, she joined the orange-Labs in Belfort-France to work on propagation modelling for UMTS and WPAN. In 1999, she joined the Broadband Wireless Access laboratory of Rennes to work on PHY layer designs based on multi-carrier techniques. She was implied in the Digitale Radio Mondiale consortium. she contributed to innovative WMAN (IEEE802.22) and short range WPANs systems based on UWB. She was involved in the IST-MAGNET and was a head of the Ultra Wideband –MultiCarrier (UWB-MC) cluster of the IST/FP6 MAGNET project. Actually, she designs UWB-OFDM WPAN systems. She concentrated her efforts on advanced interleaving algorithms for baseband processing under low cost implementation for WPAN, WLAN and mobile systems. She is actually implicated in the IST-IPHOBAC and ICT-OMEGA project She devotes some time to teaching radio communications at the University UMPC Paris VI within the SdI ESCO Master in France.